%% file: 0-main.tex
\documentclass{article}

\usepackage{fancyhdr}

\usepackage{algorithm}
\usepackage{algpseudocode}

\usepackage{listings}
\usepackage{xcolor}
\usepackage{hyperref}
\usepackage{subcaption}
\usepackage{graphicx}

\usepackage{soul,color}
\usepackage[multiple]{footmisc}
\usepackage{xurl}
\usepackage{mathtools}
\usepackage{caption}

\usepackage[a4paper]{geometry}
\newgeometry{left=1.2in,bottom=1.2in,right=1.2in,top=1.2in}

\usepackage{import}
\usepackage{multirow}
\usepackage{rotating}

\AtBeginDocument{
  \providecommand\BibTeX{{
    \normalfont B\kern-0.5em{\scshape i\kern-0.25em b}\kern-0.8em\TeX}}}

\begin{document}

\fancypagestyle{firstpage}
{
    \fancyhead{}    
    \fancyhead[L]{\textbf{Accepted by 2024 IEEE/ACM 3rd International Conference on AI Engineering - Software Engineering for AI (CAIN). DOI: \href{https://doi.org/10.1145/3644815.3644952}{10.1145/3644815.3644952}
}}
}

\title{Privacy and Copyright Protection in Generative AI: A Lifecycle Perspective}


\author{Dawen Zhang$^{1,2}$, Boming Xia$^{1}$, Yue Liu$^{1}$, Xiwei Xu$^{1}$,\\
Thong Hoang$^{1}$, Zhenchang Xing$^{1,2}$, Mark Staples$^{1}$,\\
Qinghua Lu$^{1}$, Liming Zhu$^{1}$
\\
\\
\textit{CSIRO's Data61$^1$}
\\
\textit{Australian National University$^2$}
\\
\texttt{David.Zhang@data61.csiro.au}
}
\date{}
\maketitle

\thispagestyle{firstpage}

\begin{abstract}
The advent of Generative AI has marked a significant milestone in artificial intelligence, demonstrating remarkable capabilities in generating realistic images, texts, and data patterns. However, these advancements come with heightened concerns over data privacy and copyright infringement, primarily due to the reliance on vast datasets for model training. Traditional approaches like differential privacy, machine unlearning, and data poisoning only offer fragmented solutions to these complex issues. Our paper delves into the multifaceted challenges of privacy and copyright protection within the data lifecycle. We advocate for integrated approaches that combines technical innovation with ethical foresight, holistically addressing these concerns by investigating and devising solutions that are informed by the lifecycle perspective. This work aims to catalyze a broader discussion and inspire concerted efforts towards data privacy and copyright integrity in Generative AI.
\\
\\
Keywords: Privacy, Copyrights, Generative AI, Data Lifecycle, Software Architecture, Software Engineering for AI
\end{abstract}

\maketitle

\subimport{./}{1-intro}

\subimport{./}{2-background}

\subimport{./}{3-challenges}

\subimport{./}{4-lifecycle-approach}

\subimport{./}{5-conclusion}

\input{main.bbl}

\end{document}

%% file: 1-intro.tex
\section{Introduction}
Since the dawn of the digital age, data protection, particularly concerning copyright and privacy, has been a focal point of public discourse ~\cite{regan2002privacy, litman1996revising}. The internet has enabled an unprecedented free flow and wide distribution of information on a global scale, which largely accelerated the democratization of information, fueling platforms like Wikipedia, YouTube, and StackOverflow. While this facilitated information democratization, it concurrently lowered barriers against unauthorized data use and piracy.

The success of Deep Learning (DL) owes significantly to the availability of large-scale datasets available for training DL models~\cite{sun2017revisiting}, predominantly sourced from the internet~\cite{khan2022subjects}. 
These datasets frequently encompass personal and copyrighted content~\cite{wachter2019data, khan2022subjects}, leading to a complex interplay of legal and ethical considerations.
The emergence of \textit{Generative AI} (GenAI) technologies, exemplified by applications like ChatGPT\footnote{\url{https://chat.openai.com/}}, Bard\footnote{\url{https://bard.google.com/}}, and Midjourney\footnote{\url{https://www.midjourney.com/}}, has intensified these concerns, despite marking a pivotal evolution and shift in AI applications.
GenAI models, also known as \textit{Foundation Models}, differs from traditional smaller-scale narrow-scoped DL models in its capability to generate diverse content forms, including text, audio, and visuals. 
This advancement, however, brings to the forefront the risks such as reproducing memorized data from training sets~\cite{carlini2021extracting, yang2023code}.

The legal and ethical implications of GenAI have become increasingly contentious, as evidenced by recent legal disputes (e.g., ~\cite{CopilotLawsuit, ChatGPTSuit, StableDifussionSuit}). These cases highlight the growing contention surrounding privacy and copyright concerns in AI. In a comment submitted to United States (US) Copyright Office, the US Federal Trade Commission (FTC) stated that training Generative AI models on protected expression or personal data without consent may constitute copyright infringement or privacy violation~\cite{FTCComment}.
The use of online data for AI training, commonly justified under conditions such as \textit{Legitimate Interests} for personal data and \textit{Fair Use} for copyrighted content, is now under greater scrutiny in the context of GenAI.

In response to these emerging challenges in GenAI, various approaches have been proposed to safeguard privacy and copyright. These including Differential Privacy~\cite{Bu2022}, Machine Unlearning~\cite{towards-unlearning}, and Data Poisoning~\cite{shan2023prompt,sun2022coprotector}.
While these methods offer valuable strategies, they predominantly target specific aspects and segments of the data handling process in isolation instead of the broader spectrum of challenges throughout the entire data lifecycle.

Acknowledging this critical gap, our paper aims to delve into the interconnected challenges of data protection across the GenAI data lifecycle. In pursuing this, we aim to chart new directions in AI engineering research, focusing on the development of an integrated and holistic framework. This framework is envisioned to comprehensively address data privacy and copyright protection across the entire data lifecycle, while being meticulously attuned to the intricacies of GenAI systems.

%% file: 2-background.tex
\section{Legal Basis of Privacy and Copyright Concerns over Generative AI}
Khan and Hanna~\cite{khan2022subjects} highlight two predominant legal and policy dimensions---privacy and copyright---that are pivotal in AI training contexts. To contextualize these aspects, our analysis centers on two critical legislative frameworks: the General Data Protection Regulation (GDPR) of the European Union (EU) and the Copyright Law of the United States (US). The essential provisions of these frameworks are succinctly summarized in Table~\ref{table:privacy-copyright}.

\begin{table}[h!]
\centering
\caption{The key provisions of GDPR and US Copyright Law.}
\begin{tabular}{ |c|c|c| } 
\hline
Law & Major Rights & Exemption \\
\hline
\hline
\multirow{3}{0.3\columnwidth}{GDPR (Privacy)} & Right to be Informed & \multirow{3}{0.2\columnwidth}{\centering Legitimate Interest} \\ 
& Right of Access & \\ 
& Right to Rectification & \\ 
& Right to Erasure & \\
\hline
\multirow{6}{0.3\columnwidth}{US Copyright Law (Copyright)} & Reproduce Work & \multirow{6}{0.2\columnwidth}{\centering Fair Use} \\
& Prepare Derivative Works & \\ 
& Distribute Copies & \\
& Publicly Perform & \\
& Publicly Display & \\
& Digitally Transmit & \\
\hline
\end{tabular}
\label{table:privacy-copyright}
\end{table}

\subsection{Privacy}

The GDPR encompasses several provisions that are pertinent to GenAI systems such as Large Language Models (LLMs)~\cite{zhang2023right}. Key among these are the Right to be Informed, the Right of Access, the Right to Rectification, and the Right to Erasure\footnote{Rights of the data subject: \url{https://gdpr.eu/tag/chapter-3/}}.
Under the GDPR, when personal data is collected, individuals are entitled to be informed about the collection and use of their data (Article 13, Article 14), irrespective of whether their data is acquired directly from them or through alternate sources.
They also have the right to request information about the processing of their data (Article 15), including whether or not the data is processed, the access to the personal data, the purpose and the duration of processing.

Moreover, the GDPR grants individuals the right to rectify inaccuracies in their data (Article 16) and to request the deletion of their data (Article 17). It is important to note that these rights are not absolute. For instance, the processing of personal data can be justified on the grounds of Legitimate Interest (Article 6), though these grounds may be superseded by the interests of the data subject.

\subsection{Copyright}
The US Copyright Law, as codified in Title 17 of the US Code, protects six exclusive rights\footnote{US Code Title 17 Section 106 - Exclusive rights in copyrighted works: \url{https://www.copyright.gov/title17/92chap1.html\#106}}. These rights include the rights to reproduce work, prepare derivative works based upon the work, distribute copies of the work, publicly perform the work, publicly display the work, and digitally transmit the work. It also grant authors the right to terminate licenses\footnote{US Code Title 17 Section 203 - Termination of transfers and licenses granted by the author: \url{https://www.copyright.gov/title17/92chap2.html\#203}}. The Copyright Law also introduces limitations on these rights, notably through the Fair Use doctrine (Section 107).

The application of Fair Use is determined on a case-by-case basis, and the integration of copyrighted works in AI presents new challenges distinct from traditional cases~\cite{khan2022subjects, samuelson2023generative, rodriguez2023copyright}. 
The legality of using copyrighted material in training GenAI models hinges on the specifics of each use case. Furthermore, the US Federal Trade Commission (FTC) has raised concerns in its commentary to the US Copyright Office, suggesting that using copyrighted content in AI training without consent, or commercializing GenAI outputs, could potentially constitute copyright infringement, unfair competition, or deceptive practices~\cite{FTCComment}.

In light of these legal ambiguities and the evolving nature of GenAI, there is an evident need for robust data privacy and copyright protection mechanisms. These mechanisms must be specifically tailored to address the unique challenges and stakeholder concerns in the rapidly advancing field of GenAI.

%% file: 3-challenges.tex
\section{Mapping Challenges throughout the Data Lifecycle}
\label{sec:challenges}

\begin{figure*}[]
  \centering
  \includegraphics[width=\linewidth]{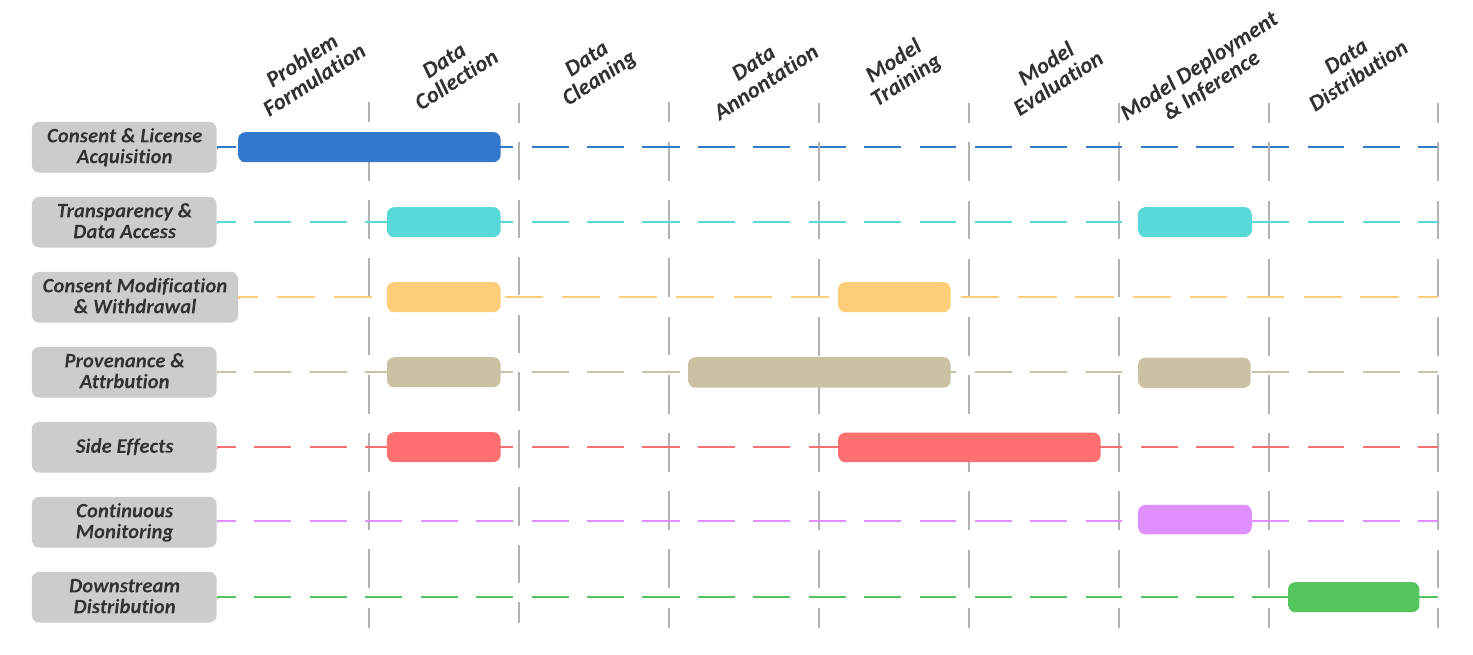}
  \caption{The key challenges throughout the data lifecycle.}
  \label{fig:mapping}
\end{figure*}

Khan and Hanna~\cite{khan2022subjects} summarized the dataset development into eight stages: problem formulation, data collection, data cleaning, data annotation, model training, model evaluation, model deployment and inference, and data distribution. In this section, the data distribution is expanded to downstream distribution to cover a broader range of distribution forms. We identify the following key challenges and maps them onto these stages of the data lifecycle (see Fig. \ref{fig:mapping}).

\subsection{Consent and License Acquisition}

The challenges of obtaining consent in GenAI are intensified by the enormous volume of data and its diverse origins.
Legal frameworks, such as the EU GDPR, mandate obtaining consent from data subjects or authors.
However, the practical implementation of these mandates is complex due to the scale and heterogeneity of data used in training GenAI models.
Novel web architectures such as SOLID~\cite{sambra2016solid} and Consent Tagging~\cite{zhang2023tag} aim to streamline consent acquisition.
SOLID, for example, is a decentralized platform designed for social web applications, enabling independent management of users' data from the applications that generate and use it. It utilizes personal online datastores that are web-accessible, granting users control over their data and the flexibility to switch between applications.
Consent Tagging empowers web users to manage their data privacy by tagging their online content, enabling better control and tracking of data usage in GenAI model training.
However, these solutions are primarily effective in scenarios where data is directly posted by individuals. They fall short in more complex contexts involving data derived from third-party sources or when the data subject is inaccessible.

A practical example of this challenge is evident in scenarios where GenAI models are trained using data from platforms like artwork sharing sites. In such cases, obtaining consent from each content creator is fraught with difficulties, including the absence of contact details, unclear permissions, and the lack of a systematic approach for managing consent in such vast and varied datasets. This not only poses logistical challenges but also raises significant ethical concerns regarding the use of such data without explicit consent.

The inability to effectively manage consent and licensing in GenAI underscores the need for more robust systems and frameworks. These systems must be capable of addressing the unique complexities of data management in GenAI, ensuring compliance with legal requirements while respecting the rights and privacy of individuals.

For the challenge of consent and license acquisition, it's closely tied to the stages of Problem Formulation and Data Collection. In Problem Formulation, defining the machine learning tasks can influence whether these tasks fall within the scopes of legitimate interest and fair use. This, in turn, affects the requirements for obtaining consent and licenses. Meanwhile, the Data Collection stage determines the sources of data, which is pivotal in identifying the origins of data that might necessitate consent. Both stages are strongly related to the challenge of consent and licensing in the data lifecycle.

\subsection{Transparency and Data Access}

The lack of transparency in GenAI training data usage poses significant challenges for data subjects and authors. A notable example is the discovery by authors that their books were utilized in training Large Language Models, often only becoming aware of this through third-party reporting. This lack of direct communication and transparency has led to legal actions, such as the class action against Meta Platforms Inc.\footnote{Chabon v. Meta Platforms Inc. (3:23-cv-04663): \url{https://www.courtlistener.com/docket/67785353/chabon-v-meta-platforms-inc/}}, highlighting the issue of using potentially pirated data in GenAI training.

In contrast to traditional data usage scenarios, where access to personal data and the visibility of copyright infringements are more straightforward, GenAI presents unique challenges. The use of personal and copyrighted data in training these models often only becomes visible through the specific outputs generated in response to certain prompts. Without in-depth investigation, identifying such uses is challenging, underscoring the need for enhanced transparency mechanisms.

Additionally, the dilemma of providing public access to datasets used in GenAI training complicates the matter. On one hand, fine-grained public access could increase awareness among data subjects and authors about the use of their data or works. On the other hand, it could exacerbate privacy and copyright concerns. This paradox mirrors the technical and ethical complexities encountered in Consent Acquisition, where balancing accessibility with privacy and copyright protection remains a significant challenge.

This challenge is closely linked with the Data Collection and Model Deployment and Inference stages. During Data Collection, it is crucial to clearly communicate about the nature and purpose of data being gathered in data collection. In the Model Deployment and Inference phase, enabling easy access for data subjects or authors to learn how their data is being used is equally important.

\subsection{Consent Modification and Withdrawal}

The GDPR endows data subjects with the right to erasure (commonly known as the ``right to be forgotten'') and the right to rectification, allowing them to demand the deletion or correction of their data.
Similarly, the Copyright Law grants authors the authority to withdraw consent for the use of their copyrighted works. These rights are pivotal in enabling individuals and creators to retain control over their data and intellectual property, even after initial sharing or utilization.

However, the practical application of these rights is significantly impeded in the GenAI domain, primarily due to transparency deficits and flawed consent acquisition processes.
This obscurity often leaves individuals oblivious to whether their personal or copyrighted data has been harvested for GenAI model training, thereby complicating the enforcement of their legal rights of consent modification and withdrawal.

Moreover, even when individuals are aware of their data's use in GenAI models and seek its withdrawal, the technical feasibility of such requests is daunting. Once data is embedded into a model's weights through training, its removal can be resource-intensive, frequently necessitating complete model retraining. Current techniques like machine unlearning have not been proven effective for large-scale foundational models.
Additionally, current guardrail approaches for blocking certain outputs are not always reliable and can be potentially exploited by simple tricks such as ``Grandma Exploit''\footnote{ChatGPT `grandma exploit' gives users free keys for Windows 11: \url{https://www.independent.co.uk/tech/chatgpt-microsoft-windows-11-grandma-exploit-b2360213.html}}.

This challenge is primarily associated with Data Collection and Model Training Stages. The origins and nature of the collected data play a critical role in the likelihood of modifications and withdrawals of consent. Furthermore, the techniques employed in Model Training may influence the feasibility and complexity of removing data from the model.

\subsection{Provenance and Attribution}
GenAI models have the capability to generate outputs that closely resemble original works which may have been used for the model training, raising significant copyright concerns\footnote{Scraping or Stealing? A Legal Reckoning Over AI Looms: \url{https://www.hollywoodreporter.com/business/business-news/ai-scraping-stealing-copyright-law-1235571501/}}, especially in instances where no license has been obtained.
Even when licenses are obtained, certain licenses mandate attribution to the original creators.
Ensuring such attribution within the outputs of GenAI models is technically challenging. The complexity of these models, coupled with the intricate manner in which they process and integrate data, makes it difficult to guarantee consistent and accurate attribution. This issue is not just a technical hurdle but also a legal one, as failure to properly attribute can lead to copyright infringement, despite the presence of a license.

Moreover, if a request for data removal is made based on the specific output, tracing the specific training data can be particularly challenging due to the inherent complexity of GenAI models that obscures the direct link between specific training data and outputs, and the lack of traceability mechanisms within these models. Such factors significantly complicate the process of identifying and removing specific data upon request, especially when the output involves fabricated or factually incorrect information generated by the model, known as hallucination~\cite{maynez2020faithfulness}. Alternatively, using band-aid solutions to filter the outputs directly is also not reliable, as mentioned previously.

The challenge of provenance and attribution arises due to complexities in Data Collection, Data Annotation, Model Training, and Model Deployment and Inference stages. Inadequate provenance and attribution often stem from the initial stages of data collection and annotation, where there may be a lack of privacy and copyright considerations. As the process progresses through model training and deployment, these challenges can become more intricate and difficult to manage.

\subsection{Side Effects}

The implementation of privacy and copyright protection measures in GenAI systems can inadvertently lead to a range of unforeseen side effects.
These may include reduced model performance~\cite{ginart2019making}, limitations on the diversity of data available for training~\cite{whang2023data}, or even potentially emergence of fairness issues if certain types of data are excluded for privacy or copyright reasons~\cite{zhang2023forgotten, koch2023no}.
Predicting the specific impacts of these protection measures on GenAI systems is challenging. While statistical models can forecast certain outcomes, such as the effects of data removal, these predictions are not always entirely accurate or comprehensive. This uncertainty necessitates a careful, ongoing evaluation process. A human-in-the-loop strategy is recommended to continuously assess and mitigate these impacts. This approach involves making informed architectural decisions and adjustments as the side effects of privacy and copyright protection measures become evident in practice.

Side effects are closely related to Data Collection, Model Training and Model Evaluation. They often emerge following data removal in response to relevant requests. Minimizing these effects necessitates a thorough understanding of the collected data, including the likelihood of data removal requests. The training methods adopted and the model evaluation also play significant roles in how these side effects manifest and can be mitigated.

\subsection{Continuous Monitoring of Privacy Breaches and Copyright Violations}

GenAI companies often deploy rudimentary safeguards such as keyword filters or prompt prefixes to mitigate privacy breaches and copyright violations\footnote{How a tiny company with few rules is making fake images go mainstream: \url{https://www.washingtonpost.com/technology/2023/03/30/midjourney-ai-image-generation-rules/}}. However, these initial measures are not foolproof and can be susceptible to circumvention~\cite{liu2023jailbreaking}.
Therefore, it is crucial to establish a mechanism of continuous monitoring.
This involves not only detecting attempts to bypass existing protective methods but also adapting and updating these methods regularly. Such vigilance is essential to ensure that privacy and copyright protections keep pace with the advancements and novel exploitation methods in GenAI.
At the same time, continuous monitoring should be complemented by a proactive approach in updating and refining protective measures.
The goal is to create a robust and adaptive system that can effectively respond to the ever-changing challenges posed by GenAI, thereby safeguarding privacy and copyright.

This challenge is particularly intertwined with the phase of Model Deployment and Inference, and this stage is essential for the prevention and detection of instances of privacy breaches or copyright violations in production.

\subsection{Downstream Distribution}

The concept of downstream distribution in the context of GenAI encompasses the transfer of personal data, copyrighted work, or derivative data to new locations, potentially extending beyond the original intended scope of use. This process poses significant legal implications, particularly in the realms of privacy and copyright law.
A critical concern in downstream distribution is the inadvertent or intentional use of these outputs for training subsequent GenAI models. Such practices can perpetuate and amplify existing privacy and copyright issues embedded within the data. This recursive cycle of data utilization and re-utilization in GenAI systems exacerbates the challenges in managing and mitigating the legal and ethical repercussions associated with downstream distribution.

Moreover, the propagation of data through downstream distribution channels complicates the traceability and accountability of data usage. It raises questions about the extent of responsibility and liability of original data providers and subsequent users, especially when the data is further processed or transformed. This scenario underscores the need for robust frameworks to manage the complexities and potential risks inherent in the downstream distribution of data within GenAI ecosystems.

While the challenges related to downstream distribution can arise from various stages of the lifecycle, they are particularly pertinent to the Data Distribution stage. The manner in which data is distributed or transferred plays a critical role in shaping the downstream privacy and copyright implications.

%% file: 4-lifecycle-approach.tex
\section{Lifecycle Approaches}

The challenges of privacy and copyright protection are pervasive throughout the entire data lifecycle. Each of these challenges can span across multiple stages, underscoring the interconnected and continuous nature of these issues. Lifecycle-centric approaches are essential for these critical issues being addressed consistently and integrally, rather than in isolation. In light of this, we highlight and call for further research into the following directions.

\noindent
\textbf{Consent across Data Lifecycle.} As demonstrated in section~\ref{sec:challenges}, issues around consent pose significant challenges in privacy and copyright protection at various stages of data lifecycle. One effort is Consent Tagging~\cite{zhang2023tag}, which uses cryptographic tags in HTTP requests and HTML DOM elements, facilitates tracking data and verifying authorship without compromising identities. However, its effectiveness is limited to scenarios where data or copyrighted work is directly posted by the data subject or owner. This limitation underscores the need for developing novel approaches that offer comprehensive consent management throughout the entire data lifecycle, addressing a wider array of data sourcing scenarios.

\noindent
\textbf{Machine Forgetting and Side Effects Mitigation.} Machine Forgetting techniques like machine unlearning, including methods like SISA~\cite{bourtoule2021machine}, can lead to side effects that necessitate mitigation strategies. These methods often require comprehensive training information and metadata from across the data lifecycle to effectively enable unlearning. This highlights the need for a holistic view of the data lifecycle and the sensible integration of machine forgetting methods.

\noindent
\textbf{Reliable Privacy and Copyright Guardrails.} It is essential to establish guardrails to protect against privacy breaches and copyright violations in GenAI. This should be not only filters on the input or output, but also a multi-layered solution throughout the entire data lifecycle. It can incorporate advanced algorithms for detecting and preventing the use of personal data or copyrighted materials, as well as sophisticated data privacy protection techniques including differential privacy training. These techniques should be an integral part of the AI system’s architecture, ensuring the reliability and robustness of the privacy and copyright protection mechanisms.

\noindent
\textbf{AI Bill of Materials for Privacy and Copyright Compliance.} Similar to Software Bill of Materials (SBOMs) in traditional software include licenses for compliance~\cite{bi2023way}, an AI Bill of Materials (AIBOMs) for GenAI should detail encrypted consent information and licenses of copyrighted works. The inclusion of this information is vital due to the significant role training data plays in GenAI, particularly regarding privacy and copyright implications. This approach necessitates a comprehensive and coordinated effort throughout the entire data lifecycle in GenAI, ensuring that the usage of data and the downstream distribution of model and derivative data are legally compliant.

%% file: 5-conclusion.tex
\section{Conclusion}
Our paper underscores the urgency to address the intertwined challenges of privacy and copyright protection from the data lifecycle perspective. These concerns, magnified by GenAI's reliance on expansive datasets, necessitate holistic approaches on top of traditional isolated methods. By mapping the challenges and advocating for solutions informed by a data lifecycle perspective, we aim to bridge the gap between technical innovation and legal responsibility. This paper contributes to the ongoing dialogue in the field, seeking to inspire collaborative efforts towards privacy and copyright protection in Generative AI.